\documentclass[11pt]{article}
\usepackage{acl}

\usepackage[colorinlistoftodos]{todonotes}
\usepackage{times}
\usepackage{latexsym}
\usepackage{graphicx}
\usepackage{booktabs}
\usepackage{float}
\usepackage{pgfplots}
\usepackage{tikz}
\usepackage{pgf-pie}
\usepackage{hyperref}
\usepackage{url} 
\pgfplotsset{compat=1.18}

\usepackage[T1]{fontenc}
\usepackage[utf8]{inputenc}

\title{Leveraging Generative AI for Enhancing Automated Assessment in Programming Education Contests}
\date{}  
\author{Stefan Dascalescu, Dumitran Adrian Marius marius.dumitran@unibuc.ro, \and Mihai Alexandru Vasiluta \\
        University of Bucharest Faculty of Mathematics and Computer Science (1,2), \\ Softbinator Technologies(2) \\ Delft University of Technology(3)}

\begin{document}
\maketitle

\begin{abstract}
Competitive programming contests play a crucial role in cultivating computational thinking and algorithmic skills among learners. However, generating comprehensive test cases to effectively assess programming solutions remains resource-intensive and challenging for educators. This paper introduces an innovative NLP-driven method leveraging generative AI (large language models) to automate the creation of high-quality test cases for competitive programming assessments. We extensively evaluated our approach on diverse datasets, including 25 years of Romanian Informatics Olympiad (OJI) data for 5th graders, recent competitions hosted on the Kilonova.ro platform, and the International Informatics Olympiad in Teams (IIOT). Our results demonstrate that AI-generated test cases substantially enhanced assessments, notably identifying previously undetected errors in 67\% of the OJI 5th grade programming problems. These improvements underscore the complementary educational value of our technique in formative assessment contexts. By openly sharing our prompts, translated datasets, and methodologies, we offer practical NLP-based tools that educators and contest organizers can readily integrate to enhance assessment quality, reduce workload, and deepen insights into learner performance.
\end{abstract}

\section{Introduction}
Competitive programming has gained substantial recognition in education for fostering computational thinking, problem-solving, and algorithmic skills \cite{wing2006computational, ackovska2015creating}. However, comprehensive and effective test creation remains labor-intensive and challenging for educators due to the need to anticipate various student solution strategies and edge cases \cite{petrovic2019automated, luxton2021assessing}. Recent advancements in Natural Language Processing (NLP) and generative AI, particularly large language models (LLMs) such as GPT-4 \cite{openai2023gpt4}, have opened new possibilities for automating complex educational tasks \cite{wang2024largelanguagemodelseducation}.

This research investigates leveraging generative NLP techniques to automatically generate robust and diverse test cases for programming problems. Our approach aims to complement expert-crafted tests, potentially reducing educators' workload and enhancing the quality of formative assessments. We specifically analyze scenarios where AI-generated tests improve upon initial expert tests, revealing additional student errors or misconceptions.

\section{Background and Related Work}
NLP techniques have increasingly been applied in educational settings to automate tasks such as automatic scoring \cite{burrows2015eras, attali2006automated}, feedback generation \cite{kochmar2020automated}, and educational data mining \cite{romero2020educational}. Generative models, in particular, have demonstrated significant potential in automating content creation and providing personalized educational experiences \cite{KASNECI2023102274}.

Previous studies have proposed methods for automated test case generation primarily using predefined templates, symbolic execution, or genetic algorithms \cite{10.1007/978-3-319-91908-9_24, fraser2011evosuite}. However, such approaches often lack flexibility or require significant domain-specific tuning. Our research differentiates itself by using generative NLP (specifically, LLMs) for dynamic, contextually appropriate test generation inspired by patterns used on competitive platforms such as Codeforces \cite{codeforces2023}.


While significant interest exists regarding large language models' (LLMs) capabilities in competitive programming contexts \cite{openai2025competitiveprogramminglargereasoning, huang2024competitionlevelproblemseffectivellm}, relatively little research has explored leveraging LLMs specifically to assist in creating problems which can be given at Olympiads and other prestigious competitive programming contests, with the only existent research to our knowledge (\cite{liu2024llmpoweredtestcasegeneration}, \cite{wang2025testevalbenchmarkinglargelanguage}, \cite{li2024largelanguagemodelstest}) involves exploring the way LLMs can help with preparing tasks given to interview coding platforms such as LeetCode, tasks that are often easier than those given at IOI style competitions. Our work directly addresses this gap by providing empirical evidence drawn from extensive historical and contemporary competitive programming datasets, impacting a broader range of problems given in contemporary contests.

Our primary contributions include introducing a novel generative NLP method for automated test case creation, empirically demonstrating its effectiveness across multiple competitive programming datasets, and openly sharing our methodology and datasets to support further research.

\section{Methodology}
\subsection{Contests Selection}
We decided to select a couple of different contests for our tests spanning different formats and platforms for our test. We focused on contests that used CMS, a widely used platform for an important set of contests where we had access to the official data, and kilonova.ro, a platform that has open access to sources and tests.

\subsubsection{OJI}
The Olimpiada Județeană de Informatică (OJI) is the county-level Computer Science Olympiad in Romania. We selected this competition due to its significance within the Romanian informatics community, backed by a long-standing tradition of over 25 years and the presence of a highly qualified scientific committee. Moreover, as presented in \cite{dumitran2024evaluatingperformancelargelanguage}, the OJI dataset has been fully translated into English and thoroughly benchmarked. Preliminary experiments using the 5th-grade problems\footnote{The OJI V problem set used can be accessed at: \url{https://kilonova.ro/problem_lists/453}} yielded promising results, motivating us to extend our evaluation by incorporating additional contests. A limitation of the OJI dataset is that we did not have access to the official submissions made during the contest; instead, we relied on the sources submitted post-contest via the Kilonova online judge. Nevertheless, the number of available submissions is substantial, making OJI one of the most resource-rich datasets for programming contests in Romania.

\subsubsection{IIOT}

The International Informatics Olympiad in Teams is an international team Olympiad in Informatics which was founded in 2016 and ever since, it became an increasingly prestigious contest in Romania and worldwide, being the only Olympiad style team contest currently held in Romania. We selected this competition due to its innovative nature, both in terms of the format as well as due to the nature of problem preparation, highly regarded as being innovative, the current team consisting of dozens of former IOI and Olympiad participants, as well as highly reputed coaches worldwide. We have obtained access to the official contest server from the organizers, which allowed us to grade the problems using the same environment and the same submissions made during the contest. In addition, a large variety of post contest source codes is available via the aforementioned Kilonova\footnote{The IIOT problem set used can be accessed at: \url{https://kilonova.ro/problem_lists/1286}} judge.

\subsubsection{Micul Gates, Info Oltenia, FII Code}
We aimed to include in our evaluation contests from 2025, as their scientific committees may have leveraged Large Language Models (LLMs) and other modern tools in the test creation process. This allowed us to investigate whether our methodology still yields consistent results under these new conditions. Consequently, we extended our experiments to recent contests hosted on the Kilonova platform. An additional advantage of using these local contests is that we had access to the official contestant submissions, providing a more complete and reliable dataset for our analysis.

FIICode\footnote{The FII Code 2025 problem set used can be accessed at: \url{https://kilonova.ro/problem_lists/1398}} is an annual programming contest held by students from UAIC, with an online qualification round and an onsite final round. The problem difficulty is usually similar to a Codeforces Div. 2 Round.

Info Oltenia\footnote{The Info Oltenia 2025 problem set used can be accessed at: \url{https://kilonova.ro/problem_lists/1342}} is an annual programming contest organized by teachers and enthusiasts from the Oltenia region in south-west of Romania. This contest has a long tradition and is essential in training the students from the region for OJI and ONI. 

Micul Gates\footnote{The Micul Gates 2025 problem set used can be accessed at: \url{https://kilonova.ro/problem_lists/1347}} is an annual junior programming contest organized in Oltenia targeted at middle school students who are starting their competitive programming journey.

These local contests, while being less prestigious than the Olympiad in Informatics, are very important for training both beginners and experts alike. Therefore, having a quality grading and testing environment in places often overlooked by problem setters is essential in order to nurture the young students' development. Thus, we found including these contests important for fulfilling the goals of our research.

\subsubsection{RoAlgo Weekly}
Furthermore, we also extened our methods to a new series of contests, RoAlgo Weekly Contests, organized by a group of volunteers from RoAlgo, the largest Romanian online competitive programming community. These contests involve very easy problems, resembling the tasks given at national informatics exams and college admission tests and we worked with the problem setting team and offered them the tools developed as part of our research. We have observed an improvement in the contest quality and the productivity of the team, as the process of preparing problems became faster, while also improving the quality of the test data.

\subsection{Platforms and Evaluation \& Reevaluation}

\subsubsection{Kilonova}
Kilonova is an open-source competitive programming platform from Romania, whose accessibility has facilitated its use in various research activities \cite{dumitran2024evaluatingperformancelargelanguage, dumitran2025exploringlargelanguagemodels}. Its open nature provides valuable features beneficial for NLP-driven research: submissions and evaluation results are publicly accessible and easy to collect programmatically; problem statements are structured in Markdown, an LLM-friendly format; and test files for most problems are conveniently downloadable. Additionally, the platform offers a straightforward API for integration and automation.

With cooperation from platform administrators, we established a mirror of the official Kilonova instance, containing a comprehensive set of historical and contemporary programming problems. Using Python scripting, we developed a semi-automated pipeline for each problem, consisting of the following steps:

\begin{enumerate}
    \item Obtaining the official model solution;
    \item Instructing the LLM to generate new test cases based on the problem statement and specified constraints;
    \item Packaging and uploading the new test cases to the mirrored platform;
    \item Selectively reevaluating previously accepted submissions to measure the effectiveness and robustness of the newly generated tests.
\end{enumerate}

This selective reevaluation capability allowed targeted assessment of the incremental value provided by AI-generated test cases without disrupting the broader user experience.

\subsubsection{CMS}

CMS (Contest Management System) has been the de facto standard online judging platform for the International Olympiad in Informatics since 2012 \cite{maggiolo2012introducing}, and is widely adopted for national and international programming contests including ICPC, OJI (since 2021), and IIOT. The widespread adoption of CMS is due primarily to its robustness, scalability, and comprehensive support for managing multiple test datasets.

In our research context, CMS provided significant advantages, notably its inherent support for parallel management of distinct sets of tests, facilitating direct and meaningful comparison of submission performance across different testing methodologies. However, the platform lacks a comprehensive API, necessitating more manual and labor-intensive processes for uploading tests and retrieving evaluation results, which somewhat limited our automation capabilities compared to Kilonova.

\subsection{Generative AI-Based Test Generation}

We utilized o3-mini-high, the newest and most powerful publicly available model developed by OpenAI for coding-related tasks. Through precise prompt engineering, we guided the model to generate tests based on patterns inspired by the Codeforces problem set. The prompts included detailed problem descriptions and explicit instructions for creating edge cases, boundary values, and complex scenarios designed to challenge diverse programming strategies. The generated tests were integrated with existing contest management systems (CMS, Kilonova) for immediate and scalable evaluation.

Leveraging our competitive programming experience, we used \textbf{testlib}\footnote{\url{https://github.com/MikeMirzayanov/testlib}}, the standard C++ library for contest tasks (used by Codeforces/Polygon). Initially, we used an LLM to generate testlib components (generator, validator, parameters, batch file). This batch file ran the generator with a model solution manually extracted from contest sources (official or Kilonova). We used English problem statements generated via prior work \cite{dumitran2024evaluatingperformancelargelanguage}.

\subsubsection{Prompting}

Our initial prompt, designed to guide the LLM in generating testing components, 
was structured as follows:
\begin{quote}
You are given a competitive programming problem in markdown. Based on this problem, please create the following tools in order to test students' source codes against a set of strong test cases.
\begin{itemize}
    \item Test case generator (ideally, you should use the \texttt{testlib.h} library developed for Codeforces). The generator should compile according to C++17 standards and you should avoid direct usages of \texttt{opt} method unless you write a function that specifically creates that template
    \item Validator for validating the tests generated
    \item Test case parameters which can be used by the testcase generator aforementioned
    \item A batch file for Windows that runs the generator for all test cases.
\end{itemize}
The test cases generated must be comprehensive, cover all possible corner cases and include tests with maximum parameters for the input constraints as well as inputs spread out (add more large tests). generate a set of 25 test case parameters which can be used by the generator. the pattern for test case names should be \texttt{test01.in}, \texttt{test02.in} etc. 
Below you get the task attached.
\end{quote}

While this initial prompt yielded promising results, we observed inconsistencies... 

Therefore, we developed an upgraded version... This revised prompt was:
\begin{quote}
You are given a competitive programming problem in markdown. Based on this problem, please create the following tools in order to test students' source codes against a set of strong test cases. 
\begin{itemize}
    \item \textbf{Test case generator:}
    \begin{itemize}
        \item It uses the \texttt{testlib.h} library developed for Codeforces
        \item The generator must be written in C++ 17
        \item Use \texttt{argvs} for parameters, \texttt{cout} for printing
    \end{itemize}
    \textit{Here is an example based on another problem which should be your model:} \\
    (here, the model code based on one of the preliminary results was attached)

    \item \textbf{Validator} for validating the tests generated
    \item \textbf{Test case parameters} which can be used by the testcase generator aforementioned

    \item \textbf{A batch file for Windows} that runs the generator for all test cases \\
    \textit{Here is a model for the batch file.} \\
    (here, the model batch file was attached)
\end{itemize}
The test cases generated must be comprehensive, cover all possible corner cases and include tests with maximum parameters for the input constraints as well as smaller tests (prioritize larger test cases). 
Generate a set of 25 test case parameters which can be used by the generator. The pattern for test case names should be \texttt{test01.in}, \texttt{test02.in} etc. 

Task is attached.
\end{quote}

This prompt enhances flexibility for complex programming challenges by making its generated parts easy to adjust.

\section{Experiments}

\subsection{Experimental Setup}

We investigated two primary applications for the LLM-generated test data:
\begin{enumerate}
\item \textbf{Complementary Role:} Augmenting existing human-authored test suites to improve coverage, potentially catching more edge cases or maximum constraint scenarios.
\item \textbf{Replacement Role:} Assessing if LLMs can fully replace human effort in test case generation for simpler problems without compromising test quality.
\end{enumerate}
All experiments used the refined prompt (detailed earlier) via the OpenAI API with English problem statements as input. For each problem, an LLM generated 25 test cases. We compared solution performance on the original human tests versus these AI-generated tests, specifically measuring how many initially 100-point solutions failed on the AI data. Findings are detailed below.

\subsection{OJI Dataset Analysis}

For the Romanian National Olympiad in Informatics (OJI) dataset, we focused on the \textbf{complementary role} (point 1 above), evaluating if LLM-generated tests could enhance existing human-curated suites.

Using an internal Kilonova instance, we replaced official tests with the 25 LLM-generated tests and re-judged previously 100-point solutions, recording the number still passing.

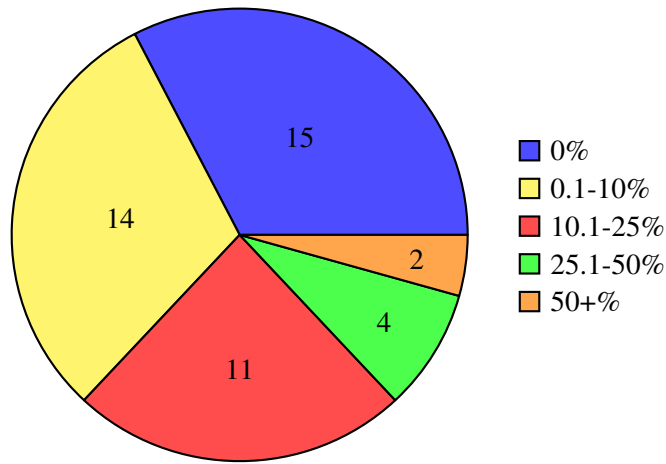
\begin{figure}[htbp]
  \centering
  \begin{tikzpicture}
    \pie[
      color       = {blue!70, yellow!70, red!70, green!70, orange!70},
      text        = legend,
      sum         = auto,
    ]{
      15/0\%,
      14/0.1-10\%,
      11/10.1-25\%,
      4/25.1-50\%,
      2/50+\%
    };
  \end{tikzpicture}
  \caption{Impact of AI-generated tests on previously accepted OJI V solutions. The chart shows the distribution of 46 problems based on the percentage of their 100-point solutions that failed when re-evaluated against the LLM-generated test set.}
  \label{fig:oji_fail_rate_distribution}
\end{figure}

Figure~\ref{fig:oji_fail_rate_distribution} shows the AI tests' effectiveness on the OJI V dataset. While $\approx$33\% (15/46) of problems showed no change for 100-point solutions, most $\approx$67\%, 31/46) saw some previously accepted solutions fail the new tests. Notably, for $\approx$13\% (6/46 problems), over 25\% of prior 100-point solutions failed (4 in the 25.1-50\% range, 2 over 50\%). This significant failure rate in a subset of problems underscores the potential for LLM-generated tests to uncover non-trivial edge cases or scenarios missed by human-authored test data, thus serving a valuable complementary role.

\subsection{IIOT}

As we got access to the official contest server, we were able to extract more complex data for the problems, thus being able to identify the number of solutions which passed both sets of test data as well as only one of them.

We have tested our method on the batch problems given at this year's preliminary rounds, the most standard category of problems given in olympiads in informatics. As we had wider access to data, we have been able to extract more information out of grading the original and the AI generated dataset.

\begin{table}[H]
    \centering
    \resizebox{0.5\textwidth}{!}{
    \begin{tabular}{l|c|c|c|c|c}
        \textbf{Problem} & 
        \textbf{100p Before} & 
        \textbf{100p After} & 
        \textbf{Both Sets} & 
        \textbf{Only Original} & 
        \textbf{Only AI} \\ 
        \hline
        walrus & 114 & 113 & 109 & 5 & 4 \\
        azugand & 49 & 49 & 47 & 2 & 2 \\
        expansionplan & 0 & 0 & 0 & 0 & 0 \\
        problemsetting & 65 & 65 & 65 & 0 & 0 \\
        binarygrid & 21 & 21 & 21 & 0 & 0 \\
        divisor & 58 & 57 & 57 & 1 & 0 \\
        homework & 0 & 7 & 0 & 0 & 7 \\
        rummy & 2 & 2 & 2 & 0 & 0 \\
        videogame & 2 & 2 & 2 & 0 & 0 \\
        tetristiling & 2 & 2 & 2 & 0 & 0 \\
        progressiveart & 56 & 43 & 41 & 15 & 2 \\
        rummy & 2 & 2 & 2 & 0 & 0 \\
        kingdomroads & 1 & 1 & 1 & 0 & 0 \\
        indexing & 81 & 51 & 51 & 30 & 0 \\
        rmi & 81 & 77 & 73 & 8 & 4 \\
        sandwich & 44 & 27 & 19 & 25 & 8 \\
        boardgame & 43 & 43 & 41 & 2 & 2 \\
        weights & 7 & 7 & 7 & 0 & 0 \\
        andqueries & 12 & 11 & 9 & 3 & 2 \\
        pali2 & 51 & 3 & 3 & 48 & 0 \\
        maxdifference & 36 & 30 & 29 & 7 & 1 \\
        lake2 & 5 & 4 & 4 & 1 & 0 \\
        pizza & 53 & 31 & 31 & 22 & 0 \\
        subjects & 117 & 104 & 103 & 14 & 1 \\
        matred & 17 & 11 & 11 & 6 & 0 \\
    \end{tabular}
    }
    \caption{IIOT results with original and AI set}
    \label{tab:iiot_results}
\end{table}

The addition of AI-generated test cases demonstrably improved the grading process for this dataset. For several problems, the AI tests proved stronger than the original human-authored ones; notably, for \texttt{pali2} and \texttt{pizza}, numerous solutions previously accepted failed the AI tests, often due to incorrect answers (WA) or exceeding time limits (TLE).

However, these results also preclude using AI tests as a complete replacement for human curation at this stage. Conversely, for problems like \texttt{sandwich}, \texttt{walrus}, and \texttt{homework}, a significant number of solutions passed the AI tests despite failing the original human-authored set, indicating the AI tests missed certain critical cases captured by the originals.

Therefore, while LLMs show significant progress in test case generation, they cannot yet reliably replace human effort entirely across all scenarios. Our findings indicate that a hybrid approach—augmenting human-curated test sets with LLM-generated cases—currently offers the most robust path toward improving test data quality and ensuring more accurate grading (i.e., maximizing the acceptance of correct solutions while rejecting incorrect ones).

\subsection{Local and Regional Contests}

Similarly to IIOT dataset, we had access to all the official submissions made by the contestants during the rounds, as well as the complete statistical data on the number of accepted solutions. However, due to the limitations of the Kilonova online judge, we were only able to test whether the new test data can help us in a complementary setup.

\subsubsection{Info Oltenia}

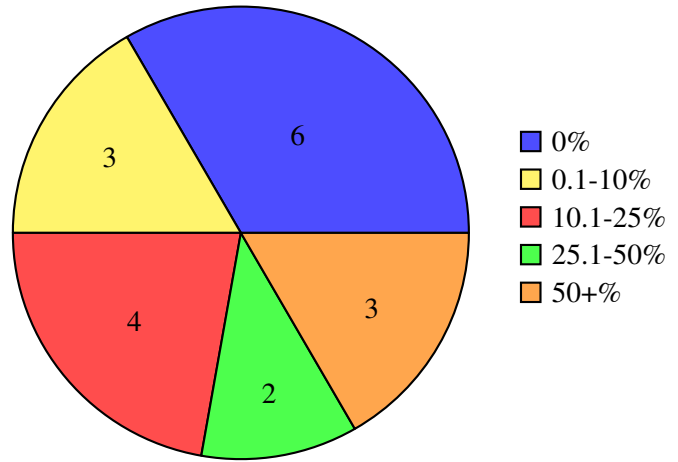
\begin{figure}[htbp]
  \centering
  \begin{tikzpicture}
    \pie[
      color       = {blue!70, yellow!70, red!70, green!70, orange!70},
      text        = legend,
      sum         = auto,
    ]{
      6/0\%,
      3/0.1-10\%,
      4/10.1-25\%,
      2/25.1-50\%,
      3/50+\%
    }
  \end{tikzpicture}
  \caption{Impact of AI-generated tests on previously accepted Info Oltenia solutions (18 problems analyzed). The chart shows the distribution based on the failure rate of 100-point solutions against LLM tests.}
  \label{fig:info_oltenia_fail_rate}
\end{figure}

Applying the same methodology to the Info Oltenia contest (hosted on Kilonova), we analyzed 18 problems, noting this contest uses distinct problem sets and committees per age group. Results are summarized in Figure~\ref{fig:info_oltenia_fail_rate}.

As shown in Figure~\ref{fig:info_oltenia_fail_rate}, the LLM-generated tests frequently identified flaws missed by the original suites. Notably, $\approx$28\% (5/18 problems) saw over 25\% of their prior 100-point solutions fail the new tests, with $\approx$17\% (3/18) exceeding a 50\% failure rate. This significant impact, potentially linked to the varied committees, highlights the value of LLM tests in complementing human-authored sets, especially where original test quality may vary.

\subsubsection{FIICode}

For the FII Code 2025 contest hosted on Kilonova, we evaluated both original and upsolved submissions against LLM-generated tests. This analysis strongly confirmed the exceptional quality of the original human-authored test cases, reflecting the contest's reputation for rigorous problem setting, often driven by Balkan/Central European Olympiad in Informatics (BOI/CEOI) and International Olympiad in Informatics (IOI) medalists.

The LLM-generated tests had a remarkably minimal impact. Across the six problems that received accepted solutions during the contest or upsolving period\footnote{Problems analyzed were \texttt{Maximize Grandi's Function} (190 AC solutions), \texttt{No More Threes} (124 AC), \texttt{Golderberg} (107 AC), \texttt{Frumusel} (89 AC), \texttt{Iggy and Bits} (41 AC), and \texttt{More or Less} (14 AC). An additional problem, \texttt{Accent}, received no accepted solutions.}, five saw absolutely no change in the verdict for submissions that initially scored 100 points when re-evaluated against the augmented test set. For the single exception, \texttt{Iggy and Bits}, where one submission out of 41 previously accepted solutions failed after the addition of the LLM tests (reducing the 100-point count from 41 to 40). This outcome, with only a single verdict change among hundreds of 100-point solutions across the contest, highlights the robustness of the original test suite and indicates limited added value from simple LLM test augmentation in this high-quality setting.

\subsubsection{Micul Gates 2025}

Applying our methodology to the Micul Gates 2025 contest on Kilonova, we found the LLM-generated tests demonstrated higher relative strength compared to the original suite for this event. Notably, no submission passed the AI tests while failing the original ones.

Furthermore, the AI tests proved strictly stronger for problem \texttt{stalpi}, where all 5 originally accepted solutions failed the new tests. For the other evaluated problems receiving accepted solutions (\texttt{joc}, \texttt{numere}, \texttt{sir}), the 100-point counts remained unchanged\footnote{Analyzed problems and initial AC counts: \texttt{joc} (28), \texttt{numere} (5), \texttt{sir} (7), \texttt{stalpi} (5). \texttt{sophie} had 0 AC.}. This suggests the LLM successfully generated more comprehensive or challenging test cases than the original set in this instance.

\subsection{Qualitative Application: RoAlgo Weekly Contests}

RoAlgo Weekly Contests are a series of contests where the challenges are of a much lower level than those given at the olympiads and programming contests, and the problem setting team has used our method to generate the test data, which has improved their work significantly as there was no need of humanly generated data anymore. The testers have checked the data generated and there were no errors whatsoever.

\section{Results Analysis and Discussion}

Our experiments across diverse datasets highlight the potential and nuances of using LLMs for test case generation in programming education contexts.

\subsection{Overall Verdict Analysis}
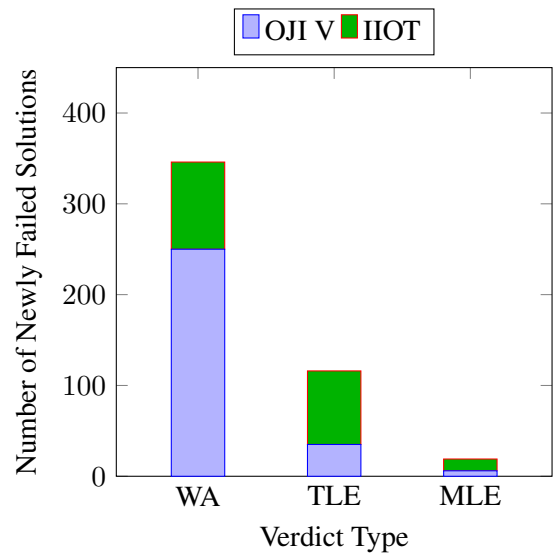
\begin{figure}[htbp]
    \centering
    \begin{tikzpicture}
        \begin{axis}[
            ybar stacked,
            width=0.95\columnwidth, 
            height=7cm,            
            symbolic x coords={WA, TLE, MLE},
            xtick=data,
            ylabel={Number of Newly Failed Solutions}, 
            xlabel={Verdict Type},                
            ymin=0,                               
            ymax=450,                             
            legend style={at={(0.5,1.03)}, anchor=south, legend columns=-1}, 
            bar width=20pt,                       
            enlarge x limits=0.3,                 
        ]
        \addplot coordinates {(WA,250) (TLE,35) (MLE,6)};
        \addplot+[fill=green!70!black] coordinates {(WA,96) (TLE,81) (MLE,13)};

        \legend{OJI V, IIOT} 
        \end{axis}
    \end{tikzpicture}
    \caption{Distribution of primary verdicts (WA, TLE, MLE) for solutions that passed original tests but failed AI-generated tests, stacked by dataset category. RE verdicts (4 instances total) omitted due to low frequency.}
    \label{fig:verdict_distribution}
\end{figure}

To understand the types of errors uncovered by the AI-generated tests across different datasets, we analyzed the distribution of verdicts for solutions that passed the original tests but failed the augmented set. The primary verdict types considered are:
\begin{itemize}
    \item WA (Wrong Answer): The program produced incorrect output on at least one test case.
    \item TLE (Time Limit Exceeded): The program failed to complete within the allocated time limit.
    \item MLE (Memory Limit Exceeded): The program consumed more memory than permitted.
    \item RE (Runtime Error): The program terminated abnormally (e.g., crash, invalid memory access).
\end{itemize}
Due to the very low frequency of Runtime Errors (only 4 instances across all analyzed datasets where AI tests caused a previously accepted solution to fail with RE), they have been omitted from the following chart (Figure~\ref{fig:verdict_distribution}) for clarity.

In the OJI V dataset, reflecting problems for younger students, Wrong Answer (WA) verdicts dominated the newly failed solutions (250 instances). This suggests the AI tests primarily caught logical errors or missed edge cases common among less experienced programmers. In contrast, the IIOT dataset, featuring more complex problems, showed a more balanced distribution between WA (96 instances) and Time Limit Exceeded (TLE) errors (81 instances), with a smaller number of Memory Limit Exceeded (MLE) cases (13 instances). This indicates the AI tests for IIOT were effective at identifying suboptimal algorithms or implementations (TLE) alongside logical flaws (WA). 

\subsection{Illustrative Cases: Successes and Challenges}
The quantitative data is complemented by specific examples. The AI tests demonstrated remarkable success in cases like \texttt{pali2} (IIOT), where 48 out of 51 previously accepted solutions failed on a maximal TLE-inducing test missed by the original setters. Similarly, for \texttt{cartele} (OJI), nearly two-thirds of solutions failed on various corner cases identified by the AI. This often occurred with older problems where manual test generation standards might have been less rigorous, highlighting the AI's ability to systematically explore edge conditions.

However, the LLM approach faced challenges with certain problem types, particularly those involving complex geometric properties or very specific input constraints, such as \texttt{The Dutch Farmer} (IIOT) and \texttt{Vedere} (InfoOltenia). Generating valid and meaningful tests for such problems remains difficult even for humans and represents an area requiring more sophisticated prompting or validation. Furthermore, as seen in the IIOT analysis (e.g., \texttt{sandwich}, \texttt{walrus}), AI tests sometimes missed cases caught by human experts, leading to solutions erroneously passing the AI set.

\subsection{Implications for Assessment and Education}
Our findings strongly support the use of LLM-generated test cases in a \textbf{complementary role}. They demonstrably enhance existing test suites by uncovering errors missed by human setters, particularly for edge cases and performance limitations. This directly improves the accuracy and fairness of assessments. As evidenced by the RoAlgo Weekly contests, this approach can also increase the productivity of problem-setting teams, especially for less complex problems, by providing a strong baseline set of tests.

The results also clearly indicate that current LLM-based generation is not yet reliable enough for \textbf{full replacement} of human-authored tests in all scenarios, especially for complex problems or high-stakes competitions. The instances where AI tests were weaker than human tests highlight the need for expert oversight.

The most effective approach appears to be a \textbf{hybrid model}: leveraging LLMs to generate a broad set of candidate tests, including challenging boundary and performance cases, followed by human expert review, selection, and potential augmentation. This combines the scalability and systematic exploration of AI with the nuanced understanding and validation capabilities of human experts.

Furthermore, integrating AI-generated tests can provide valuable formative feedback, helping educators identify common student misconceptions or areas where algorithmic understanding is weak (e.g., distinguishing WA-prone vs. TLE-prone problems). Reducing the burden of manual test creation can free up educator time for more direct student interaction and instructional design.

\section{Future Work}
\label{sec:future}
While initial results are promising, we can significantly improve outcomes for certain problems by using more specific prompts for the generator, such as instructing models to output code for specific graph types.

Additionally, experimenting with more LLMs beyond OpenAI's o3-mini-high could provide valuable comparisons of different generation methods. We also note that generating more than the current 25 test cases per problem would better align with real-world competitive programming requirements, especially for difficult problems.

Building on this, we propose three research directions:

\begin{itemize}
\item \textbf{ICPC-Style Contests}: Adapt the methodology for team competitions.
\item \textbf{Platform Generalization}: Validate on more platforms (e.g., LeetCode, HackerRank, university systems, other olympiads).
\item \textbf{Human-AI Co-Design}: Develop tools for educator-guided refinement and AI-suggested edge cases for human validation.
\end{itemize}

These directions aim to test the limits of automated generation while ensuring alignment with real-world assessment practices.

\section{Limitations}
\label{sec:limitations}

While our approach demonstrates significant promise in automating test generation for programming contests, several limitations merit discussion:

\begin{itemize}
    \item \textbf{Platform Coverage:} Our analysis focused primarily on contests hosted on the Kilonova.ro platform and the IIOT dataset (evaluated using CMS). While these represent diverse formats (national olympiads, team competitions, online platforms), they do not encompass all important paradigms like ICPC-style contests or widely used platforms such as Codeforces or AtCoder. Expanding to these platforms could reveal context-dependent variations in test-generation efficacy but faces challenges in accessing both contestant solutions and original test cases due to privacy and intellectual property constraints.

    \item \textbf{Model Dependencies:} The quality and effectiveness of the generated tests are intrinsically linked to the capabilities of the underlying LLM (in our case, OpenAI's \texttt{o3-mini-high} model\footnote{You might want to specify if this is known to be based on GPT-4 or a similar architecture, if permissible.}) and the precision of the prompt engineering. Performance may vary significantly when using different LLMs (e.g., open-source models or those from other providers) or less optimized prompts. While we release our final prompts to aid reproducibility (\textit{see Appendix X} - *add appendix reference if applicable*), the core model capability remains a key factor.

    \item \textbf{Generalizability for Full Replacement:} Our findings strongly support the use of LLM-generated tests in a \textit{complementary} role to enhance existing human-authored suites, particularly effective for identifying edge cases or performance issues missed in older or less rigorously tested problem sets (e.g., OJI V, Info Oltenia). However, the results, particularly from the high-quality FIICode contest and instances in the IIOT dataset where AI tests missed errors caught by human tests, indicate that current LLM-based generation is not yet consistently reliable enough for \textit{full replacement} of expert-curated tests, especially in high-stakes competitions or for problems with very complex logical or constraint structures. Human oversight and validation remain essential.

    \item \textbf{Cost and Scalability:} Although utilizing proprietary LLM APIs can raise concerns about operational costs, our extensive evaluation across multiple contests demonstrated exceptional cost-effectiveness. The entire process of generating 25 test cases for each analyzed problem incurred a total API cost of only \textbf{\$4.64 USD}. This was achieved through an efficient combination of targeted API calls (averaging approximately \textbf{\$0.1 USD per problem}) and leveraging free user interface interactions during development where feasible. This low cost underscores the method's practicality and affordability for educators and contest organizers seeking substantial improvements in test coverage and potential time savings compared to manual creation, without significant financial investment.

    \item \textbf{Fixed Number of Generated Tests:} We standardized on generating 25 test cases per problem for this study. While effective in revealing previously undetected errors across various datasets, this fixed number may not be universally optimal. Real-world competitive programming practices often involve larger test sets, especially for more difficult problems. Future work could investigate generating a larger or adaptive number of tests based on problem complexity or type, although this would proportionally impact the (currently very low) generation cost.
\end{itemize}

\section{Ethical Considerations}
\label{sec:ethics}

The use of generative AI in educational assessments raises several ethical concerns that require careful mitigation:

\begin{itemize}
    
    \item \textbf{Transparency}: All AI-generated content in our experiments is clearly documented, with prompts and methodologies openly released to enable scrutiny \cite{mitchell2019model}.
    
    \item \textbf{Data Privacy}: Contestant solutions were anonymized and used in compliance with GDPR and platform terms of service. No personally identifiable information was processed by our models.
    In fact the contestant data was never given to the models and only the open available problem definition were offered to them
\end{itemize}


\bibliography{main}

\appendix

\section{Appendices}
\label{sec:appendix}

Longest Palindrome abridged statement: You are given a sequence of $N$ positive integers on the blackboard, where some of the numbers have been erased and replaced with $-1$. Now, we want to restore the sequence by replacing the $-1$ values with the same number of his choice. Your task is to determine the length of the longest palindromic contiguous substring that can be obtained after choosing an optimal value for $x$.

TLE test for Longest Palindrome: $N = 200000$, $a_1 = a_2 = \dots = a_n = -1$

Cartele abridged statement: You are given an access card system developed in a school, with every student having one such card. The system prints every day the log of the students, with various information shown. Knowing the set of information, find the number of boys and girls who are still at school, the number of seconds where we had at least one student in school and the biggest timespan where an odd number of boys were at school at the same time.

WA test for Cartele:

$C = 3$, $N = 8$, $logs = [[b \ i \ 0 \ 10 \ 28], [f \ i \ 0 \ 10 \ 30], [b \ e \ 0 \ 10 \ 33], [f \ e \ 0 \ 10 \ 40]$ \\, $[b \ i \ 0 \ 10 \ 41], [f \ e \ 0  \ 10 \ 48], [f \ i \ 0 \ 10 \ 58], [f \ i \ 0 \ 11 \ 4]]$

\end{document}